\documentclass[twocolumn,superscriptaddress,showpacs,amsmath,amssymb]{revtex4}
\usepackage{latexsym, amscd}
\usepackage{amsmath}
\usepackage{graphicx}

\renewcommand{\l}{\left}
\renewcommand{\r}{\right}

\newcommand{\e}[2][-\imath]{e^{#1 #2}}
\newcommand{\ii}{\imath}
\newcommand{\shalf}{\frac{1}{\sqrt{2}}}
\newcommand{\half}{\frac{1}{2}}

\newcommand{\f}[1]{\textrm{#1}}
\newcommand{\B}[1]{\mathbf{#1}}
\newcommand{\eq}[1]{\begin{equation}#1\end{equation}}

\newcommand{\eqa}[1]{ \begin{eqnarray}#1\end{eqnarray}}
\renewcommand{\vector}[4][r]{
\left(\begin{array}{#1}
#2 \\
#3 \\
#4 \\
\end{array}\right)}

\begin{document}

\title{Oscillating solitons in a three component BEC}

\author{Piotr Szankowski}
\affiliation{Institute for Theoretical Physics, Warsaw University,
ul. Ho\.{z}a 69, PL-00-681, Warsaw, Poland}
\author{Marek Trippenbach}
\affiliation{Institute for Theoretical Physics, Warsaw University,
ul. Ho\.{z}a 69, PL-00-681, Warsaw, Poland}
\author{Eryk Infeld}
\affiliation{Soltan Institute for Nuclear Studies, ul. Ho\.{z}a 69, PL-00-681 Warsaw, Poland}
\author{George Rowlands}
\affiliation{Department of Physics, University of Warwick, Coventry CV4 7AL, England}

\begin{abstract}
We investigate the properties of three component BEC systems with spin exchange interactions. We consider different coupling constants from those very special ones leading to exact solutions known in the literature. When two solitons collide, a spin component oscillation of the two emerging entities is observed. This behavior seems to be generic. A mathematical model is derived for the emerging solitons. It describes the new oscillatory phenomenon extremely well. Surprisingly, the model is in fact an exact solution to the initial equations. This comes as a bonus.
\end{abstract}

\maketitle

\date{\today}

The idea of spinor condensates was first suggested in seminal papers of Ho and Ohmi \cite{intro:ref}. The experimental creation of spinor condensates \cite{intro:spinor1}, in which the spin degree of freedom, frozen in magnetic traps, comes into play, opened a new perspective to various phenomena that are not present in single component Bose Einstein condensates. These included the formation of spin domains \cite{intro:spinor2} and spin textures \cite{intro:spinor3}. A theoretical description of the formation of spin domains can be found in \cite{ref3theory}. A spinor condensate formed by atoms with spin $F$ is described by a macroscopic wave function with $2F+1$ components. Here we focus on the $F=1$ case, which has been studied in a number of works \cite{ref3,ref4}. Multicomponent vector solitons with $F=1$ have been predicted; bright in \cite{bright:spinor:soliton} and dark in \cite{dark:spinor:soliton}.

In this Letter we investigate the dynamics of an F = 1 spinor Bose Einstein condensate for a wide range of scattering length. In particular we address the general physical problem of spin soliton collisions. For one specific ratio of the scattering lengths, Wadati and coworkers in a series of papers \cite{Wadati:ref} found a complete classification of the one soliton solution with respect to the spin states. They even presented an explicit formula of the two-soliton solution. One soliton solutions come in two classes: polar (spinless) solitons, and ferromagnetic solitons \cite{Wadati:ref,future}. Both can be generalized to a wider set of scattering lengths. In this Letter we focus on one of these classes - non-ferromagnetic (polar) solitons and analyze their collisions. As a result of these collisions we obtain new entities. These are solitons with populations oscillating between different spin components. Although the total spin is conserved, each emerging soliton carries nonzero spin. When we look closer we are able to describe these emerging entities by a model analytical expression.

To begin with, we consider a dilute gas of trapped bosonic atoms with hyperfine spin $F = 1$. The wavefunction in vector form is $\B\Psi\l(x,t\r)=\l(\Psi_1,\Psi_0,\Psi_{-1}\r)^T$ and it must satisfy the spinor Gross Pitaevski equation
\eqa{\label{mainequation}
\ii\hbar\partial_t\B{\Psi}
&=&
\l[ -\frac{\hbar^2}{2 M}\partial_x^2
 + c_0\B{\Psi}^\dagger \cdot\B{\Psi} +\r. \nonumber\\
&&\l.+ c_2\l(
	 \sum_{\alpha=1}^3\l(\B{\Psi}^\dagger\cdot \hat f^\alpha \cdot\B{\Psi}\r)\hat f^\alpha
\r)\r]\B{\Psi}.
}
Here the $\hat f^\alpha, (\alpha=1,2,3)$ are angular momentum operators in a 3x3 representation
and $c_0$ is negative to allow for bright soliton formation. We choose a natural set of scales characteristic for the problem: length scale - $l = \frac{\hbar^2}{2 M\left|c_0\right|}$ and time scale $\bar{t} = \frac{\hbar l}{|c_0|}$.  When we express Eq.~(\ref{mainequation}) in these units, divide the equation by $c_0$, all the coefficients but the ratio between self and cross nonlinear coupling $\frac{-c_2}{|c_0|}$, which we denote by $\gamma$, will be equal to one.

%
%

In the paper of Ieda {\it et al} \cite{Wadati:ref} the authors considered this system with coupling constants $c_2=c_0\equiv c <0$ ($\gamma=1$). In this case Eq.~(\ref{mainequation}) describes a completely integrable system. The authors find N soliton solutions via the Hirota method \cite{Wadati:ref}. In particular, they present both $\f{N}=1$ and $\f{N}=2$ solutions explicitly.

Another special case occurs when $\gamma=0$ and different spin components are separated. The two above cases are the only ones for which $N$ soliton solutions have been found. To our surprise, when $\gamma$ differs from zero and one, as a result of colliding two $\f{N}=1$ solitons a new class of stable, localized but oscillating entities emerged. The details of the numerical and analytical studies are presented below.

Consider the polar one soliton solution to Eq.~(\ref{mainequation})
\eq{\label{onesoliton}
\B\Psi = \sqrt{2}\, k\, \f{sech} \left[ k\left(x - x_0 - 2p t\right)\right]\e[\ii]{p x}\,\e[\ii]{(k^2-p^2)t}\,\B\chi,
}
where
\eq{
\B\chi=
\vector[c]
{-\shalf \e{(\beta_i-\tau_i)}\sin\alpha_i}
{\e[\ii]{\tau_i}\cos\alpha_i}
{\shalf \e[\ii]{(\beta_i+\tau_i)}\sin\alpha_i}
 = \e[\ii]{\tau_i}\hat{\cal U}_i \vector{0}{1}{0}\nonumber
}
is parameterized by three angles $\tau_i,\alpha_i,\beta_i$. The rotation operator used here is $\hat{\cal U}_i \equiv \hat{\cal U}(\beta_i,\alpha_i,\theta_i) =\e{\hat f^3\beta_i}\e{\hat f^2\alpha_i}\e{\hat f^3\theta_i}$. Note that $\theta_i$ is in this case a dummy variable; our spinor $(0,1,0)^T$ is not affected by $\e{\hat f^3\theta_i}$.  Equation (\ref{onesoliton}) describes a stationary modulus soliton moving with velocity $2p$. The norm (or total number of particles) of this soliton is: $N_T=\int\f{d}x\,n\l(x,t\r)=\int\f{d}x\,\B\Psi^\dagger\cdot\B\Psi = 4k$, total momentum: $P_T=\int\f{d}x\,p\l(x,t\r)=\int\f{d}x\,\B\Psi^\dagger\cdot\l(-\imath\partial_x\r)\B\Psi=p\,N_T$ and the spin density: $f^\alpha\l(x,t\r)=\B\Psi^\dagger\cdot\hat f^\alpha\cdot\B\Psi=0$. This solution is valid for arbitrary $\gamma$.

%
%
%

\begin{figure}[tbp]
\includegraphics[width=8.5cm]{grid%
.eps}
\caption{Top: Density plot of $2|\Psi_{\pm 1}|^2$ (left) and $|\Psi_{0}|^2$ (right) for two soliton collision. Here $\gamma=-1$, $k_1=k_2=1/4$, $p_1=0$, $p_2=1.65$, $\tau_1=\tau_2=\beta_1=\beta_2=0$, $\alpha_1=0$ and $\alpha_2=\pi/4$. \\ Bottom: Details of the initially stationary soliton pictured after the collision and to a different (much longer) time scale. Observe the oscillatory character of the soliton components and small momentum transfer (panels c and d).} \label{fig1}
\end{figure}

\begin{figure}[tbp]
\includegraphics[scale=0.45]{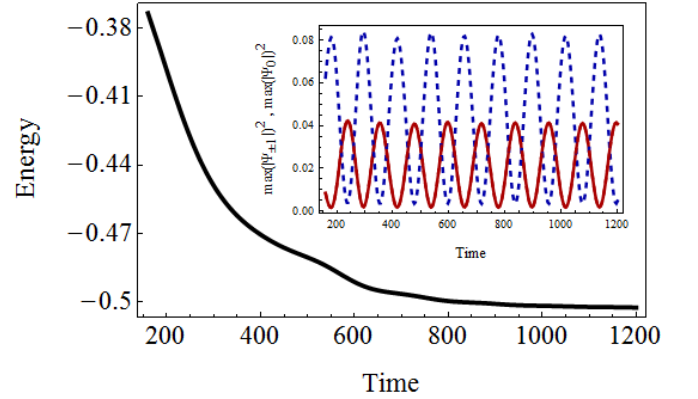}
\caption{(Color online) Total energy of the soliton on the left in Fig.1 as a function of time. Energy is lost to radiation. Energy in units of modulus of the energy of the initial soliton. Inset: peeks of modulus of $|\Psi_0 |^2$ (dashed blue line) and $|\Psi_{\pm 1} |^2$ (solid red line) of the emerging soliton. Observe the harmonic oscillations.}\label{fig2}
\end{figure}

\begin{figure}[tbp]
\includegraphics[scale=0.4]{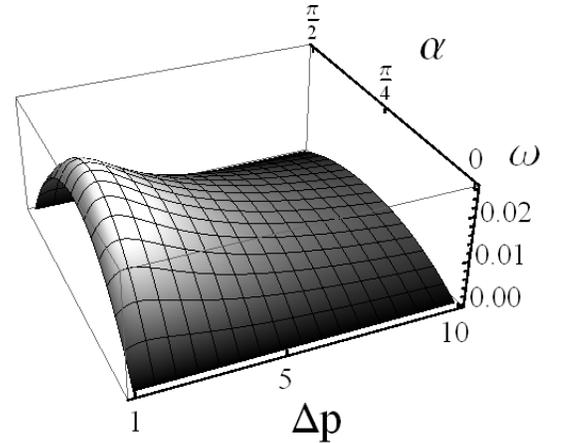}
\caption{Frequency of the oscillatons after the collision as a function of the collision parameter $\alpha_i$ and relative momentum $\Delta p$. Note that there is a maximum for small momentum and $\alpha=\pi/4$. The sector ($\pi/2,\pi$) is a mirror image of this graph. Here $\gamma=-1$. } \label{fig3}
\end{figure}


We collide two solitons of the type described by Eq.~(\ref{onesoliton}), with equal norms $k$, but otherwise different parameters. Due to the Galilean and rotational invariance we can set one of the solitons to be static with $\chi=(0,1,0)^T$. The second soliton will move with momentum $\Delta p$ and general $\alpha_i$, $\beta_i$ and $\tau_i$ angles. In the case of $\gamma=1$, two colliding solitons of the kind described in Eq.~\ref{onesoliton} indeed emerge in the same form. Below we show that for different configurations (values of $\gamma$) a new kind of solution can be observed after the collisions, and after an interval of radiation (Figs. \ref{fig1} and \ref{fig2}). After the collision is completed, we observe a pair of solitons that do not have a product structure. Spin and spatial degrees of freedom cannot be separated. The population oscillates between different spin components. We observe a well defined oscillation frequency. We will refer to these new solutions as oscillatons.

The result of the collision depends on the set of angles ($\alpha_i, \beta_i, \tau_i)$ and relative momentum $\Delta p$. In particular, we observe that the frequency of oscillations depends only on $\alpha_i$ and $\Delta p$. The most striking feature of the collision is the appearance of a local spin density; the colliding partners carry away equal and oppositely polarized spin densities. Moreover the spin vectors will lie in the $XY$ plane. The initial angle $\beta_i$ determines the the final orientation, but not the frequency! This was verified in simulations. Figure \ref{fig3} presents a three dimensional surface of the frequency of oscillations after the collision is completed, versus $\alpha_i$ and relative momentum $\Delta p$. The maximum frequency occurs at $\alpha_i=\pi/4$. For $\alpha_i=0$ we recover the $\gamma=1$ case; collisions are elastic. More interestingly, for $\alpha_i=\pi/2$ there is practically no interaction between colliding solitons. The oscillation frequency is also a function of relative momentum. We observe a clear maximum for small values of $\Delta p$. Also for small relative momentum we see in the numerical simulations that some energy is transferred to the external degrees of freedom. In our case the initially static soliton is set in slow motion (see Fig.~\ref{fig1}c and \ref{fig1}d).

We describe the collision in the reference frame of one of the initial solitons. The stability of the static wavepacket was investigated numerically.  During the collision and for some time afterwards, part of its energy was lost due to radiation. Asymptotically, the energy of the stationary solution converged to a constant value, which was larger that the energy of the initial soliton (see Fig.~(\ref{fig2})).


In what follows we propose an analytic solution to Eq.~(\ref{mainequation}) that fits the emerging oscillatons and turns out to be exact. We focus on the stationary soliton, since the moving partner is essentially described by the same wavefunction, but with a Galilean boost. We assume the following form of solution, hoping to describe the emerging oscillatons
\eq{\label{Ansatz}
\Psi=
 \e[\ii]{\tau_f}\hat{\cal U}_f
 \l[
  \eta_+(x)\e[\ii]{\mu_+t}\vector{1}{0}{0}
 -\eta_-(x)\e[\ii]{\mu_-t}\vector{0}{0}{1}
 \r],
}
where $\hat{\cal U}_f=\hat{\cal U}(\beta_f,\alpha_f,\theta_f)$. This ansatz, by its very form, describes oscillations between components. This may be seen by looking at the squares of the modulai
{\setlength\arraycolsep{0.1em}
\eqa{
\l|\Psi_0\r|^2 &=&
 \half\sin^2\alpha_f\Big(\eta^2_++\eta^2_- 
+2\eta_+\eta_-\cos\l[\omega t-2\theta_f\r]\Big)\label{eq5}
}
\eqa{
\l|\Psi_{\pm 1}\r|^2&=&
 \eta^2_\pm\sin^4\frac{\alpha_f}{2} +
 \eta^2_\mp\cos^4\frac{\alpha_f}{2} +\nonumber\\
&&-\half\eta_+\eta_-\sin^2\alpha_f\cos\l[\omega t-2\theta_f\r] \label{eq4}
}
}
All the spin components oscillate with frequency $\,\omega=\mu_+-\mu_-$ \cite{future}. The total global spin for this solution is a nonzero constant! Notice that $\alpha_f$ is related to the amplitude of the oscillations and by examining the spin density
\eq{
\B{f}=\B{e_\alpha}(\B{\Psi}^\dagger\cdot\hat f^\alpha\cdot\B{\Psi})=\l(\eta_+^2-\eta_-^2\r)\B{n},
}
where $\B n=(\sin\alpha_f\cos\beta_f,\sin\alpha_f\sin\beta_f,\cos\alpha_f)^T$, we find that $\beta_f$ influences the spin vector orientation. We see that $\theta_f$ only plays the role of a time delay. We also confirmed that the modulus of the local spin, $\l|\B{f}\r|=\l|\eta^2_+-\eta^2_-\r|$, and the density $\B\Psi^\dagger\B\Psi=\eta^2_++\eta^2_-$ are time independent.

Our anzatz substituted into Eq.~(\ref{mainequation}) leads to the following system of two coupled {\it ordinary} differential equations
\eq{
-\mu_\pm+(1+\gamma)\eta^2_\pm+(1-\gamma)\eta^2_\mp+\frac{{\eta_\pm}''}{\eta_\pm}=0. \label{eqneta}
}
The problem has thus been reduced to solving ordinary differential equations, a much simpler task than solving Eq.~(\ref{mainequation}). Assuming that this can be done, our ansatz can be considered as an exact solution.

\begin{figure}[tbp]
\includegraphics[scale=0.45]{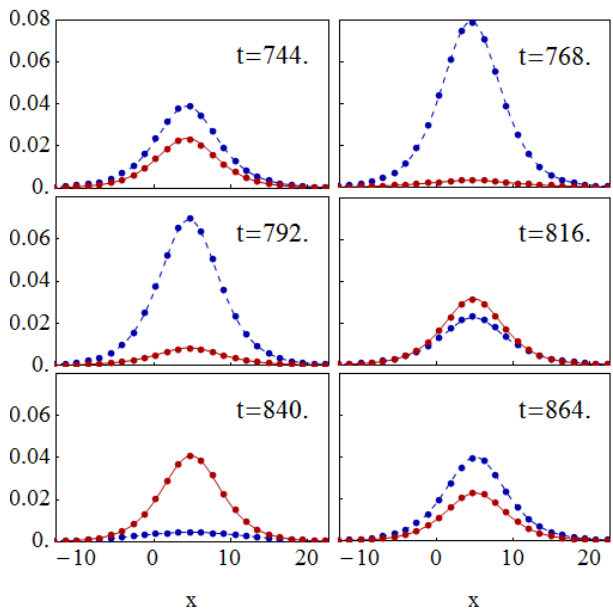}
\caption{(Color online) Comparison of the predictions of our analytical model with an emerging oscillaton found numerically. Collision parameters are as in Fig.~\ref{fig1}. We have adjusted $\mu_+$, $\omega=\mu_+-\mu_-$ and the angles $\alpha_f,\beta_f,\theta_f,\tau_f$ in Eq.~(\ref{Ansatz}) to fit the simulation. There is perfect agreement between the red (continuous) line corresponding to the analytic solution for $|\Psi_{\pm 1}|^2$ given in Eq.~(\ref{eq4}) and the red dots obtained in the numerical simulation. Blue (dashed) lines and dots correspond to $|\Psi_0|^2$ (Eq.~(\ref{eq5})).}\label{fig4}
\end{figure}

We discovered that the results of the numerical studies reported above can be well described by our ansatz. For example, Fig. (\ref{fig4}) compares predictions of our model with an emerging oscillaton found numerically. The match is perfect! The adjustment requires finding $\mu_+$, $\mu_-$ and the set of angles $\{\beta_f, \alpha_f, \theta_f\}$, which is rather straightforward, and solving the coupled equations (\ref{eqneta}).

The new solution (\ref{Ansatz}) can be analyzed independently of the context of soliton collisions, as a new class of oscillaton solutions to Eq.~(\ref{mainequation}). A particular case is when $\gamma=1$, for which a solution for the pair of equations (\ref{eqneta}) is obtained explicitly. For this $\gamma$ the equations for $\eta_\pm$ are independent and each can be solved analytically.
The solutions of interest are
\eq{
\eta_\pm\l(x\r)=\sqrt{\mu_\pm}\,\f{sech}\l(\sqrt{\mu_\pm}\,x\r).
}
To understand the significance of this result one should go back to Eq.~(\ref{Ansatz}) and appreciate its complex character. We found a solution that looks like being composed of two solitons! It oscillates with frequency $\omega = \mu_+-\mu_-$.

For general $\gamma$, we have developed methods for dealing with two distinct regions. The first considers $\eta_+$ and $\eta_-$ of the same magnitude, the second covers $\eta_+ \gg \eta_-$.

When $\mu_+=\mu_-$ we have no oscillations ($\omega=0$). This is equivalent to the solution given by Eq.~(\ref{onesoliton}). In this case $\eta_+=\eta_-=\sqrt{\mu}\,\f{sech}\l(\sqrt{\mu}\,x\r)$. When $\mu_+$ and $\mu_-$ are {\it almost} equal $\mu_\pm=\mu (1\pm \delta)$ we have found an approximate solution
\eq{
\eta_\pm\l(x\r)=\alpha_\pm\,\f{sech}^{\sqrt{\mu_\pm/\mu}}\l(\sqrt{\mu}\,x\r).
}
When we feed this ansatz into Eq.~(\ref{eqneta}) we obtain two approximate equations for the amplitude
\eq{
\mu_\pm+\sqrt{\mu\,\mu_\pm} \thickapprox (1+\gamma)\alpha^2_\pm+(1-\gamma)\alpha^2_\mp.
}
These equations yield the values of the amplitudes $\alpha_\pm$ and thus conclude our search for an approximate solution. We have checked this solution against numerics for values of $\gamma$ in the interval $[-1/2,1/2]$ and $\delta$ between 0.05 and 0.1. In all cases the fits were very close.

On the other hand, consider the case when one of the etas is much larger than the other, say $\eta_+ \gg \eta_-$, any $\gamma$ (excluding $\pm 1$). If we ignore $\eta_-$ in the first equation, we obtain for $\eta_+=\sqrt{2\mu_+/(1+\gamma)}\,\f{sech}\l(\sqrt{\mu_+}\,x\r)$. We next insert this value into the equation for $\eta_-$, where it plays the role of a potential:
\eq{
{\eta_-}''+\l[-\mu_-+2\mu_+\l(\frac{1-\gamma}{1+\gamma}\r)\f{sech}^2\l(\sqrt{\mu_+}\,x\r)\r]\eta_-=0. \label{eqnetapo}
}
Fortunately, this equation can be solved. The solution is
\eq{\eta_- \propto \f{sech}^{\epsilon}\l(\sqrt{\mu_+}\,x\r)
P^{(\epsilon,\epsilon)}_{2n}\l[\f{tanh}\l(\sqrt{\mu_+}\,x\r)\r],
}
where $\epsilon=\sqrt{\mu_-/\mu_+}$, $P^{(a,b)}_{m}[\xi]$ is a Jacobi polynomial which must be finite of order $n$. This demands that $\mu_-$ be one of the values given by
\eq{
\mu_-=\frac{\mu_+}{4}\l(-1-4n+\sqrt{1+8\frac{1-\gamma}{1+\gamma}}\r)^2,
}
a spectrum. Our conclusion is that the values of $\mu_-$ come in a discrete spectrum.

In conclusion, we considered three component BEC systems with spin exchange interactions and a general coupling constant $c_2$ and any negative $c_0$. We focused on two polar soliton inelastic collisions and discovered that robust oscillating entities are generated. Shortly after the collision, some of the energy is radiated away, but the emerging entities are again solitons, with populations oscillating between spinor components with a well defined frequency and nonzero spin density. This frequency depends on the collision parameters. We derived a mathematical model which fits the emerging solitons perfectly. Surprisingly, the model is an exact solution to the initial Gross Pitaevskii equations. This solution is not limited to the simple product form of the spatially dependent and spinor parts.

Experimental realization of the ideas presented here is perfectly feasible. It could be planned along the lines of the work concerning matter wave soliton trains \cite{Strecker}, using Lithium. At first, a polar wavepacket should be produced in a tight optical trap, e.g. radial frequency 300 Hz and longitudinal frequency 100 Hz. Next, a magnetic field of 700 G of constant magnitude should be applied to change the scattering length via Feshbach resonance to a slightly positive value, say 0.1 nm, and the longitudinal trapping frequency should be gently decreased to 10 Hz, to generate an elongated structure. A pair of Bragg pulses can be applied to generate counterpropagating solitons. These solitons will travel towards the edges of the trap and be rebounded there. Before they collide, a specially prepared magnetic field, perpendicular to the original one, should tilt the relative spin direction. After the collision, which is inelastic, we measure the components of the oscillatons.  We numerically simulated this very sequence and confirmed the result. It was as expected.

Bright solitons such as considered here exist only for attractive interactions. When interactions are repulsive, solitons, if they appear, are dark. They will be considered separately.

The authors acknowledge support of a Polish Government Research Grant and also the Foundation for Polish Science Team Programme co-financed by the EU European Regional Development Fund.

\end{document}